\documentclass{PoS}

\newcommand{\e}{\epsilon}
\newcommand{\beq}{\begin{eqnarray}}
\newcommand{\eeq}{\end{eqnarray}}
\newcommand{\secn}[1]{Section~\ref{#1}}
\newcommand{\npo}{{n+1}}
\newcommand{\pn}{\Phi_n}
\newcommand{\pnpo}{\Phi_\npo}
\newcommand{\NLO}{{\mbox{\tiny{NLO}}}}
\newcommand{\dd}{{}}
\newcommand{\poh}{\hat{\Phi}_1}
\newcommand{\fd}{{(4)}}
\newcommand{\po}{\Phi_1}
\def\eq#1{Eq.~(\ref{#1})}
\def\bra#1{%
  \left\langle\smash{#1}{\vphantom1}\right|}
\def\ket#1{%
  \left|\smash{#1}{\vphantom1}\right\rangle}
\def\slash#1{#1 \hskip-0.45em /}

\title{Factorization and subtraction}

\ShortTitle{Factorisation and subtraction}

\author{Lorenzo Magnea\footnote{Speaker}\, , \,
            Ezio Maina, \,
            Paolo Torrielli, \,
            Sandro Uccirati\\
      University of Torino and INFN, Sezione di Torino\\
      E-mail: \email{lorenzo.magnea@unito.it}}

\abstract{We explore the connection between the factorisation of virtual 
corrections to multi-particle massless gauge theory amplitudes and the problem 
of subtraction at NNLO and beyond. Taking inspiration from virtual factorisation, 
we provide a set of definitions for local soft and collinear counterterms, expressed
in terms of matrix elements of operators involving fields and Wilson lines, and 
valid to all orders in perturbation theory. We hope that the connection between 
factorisation and subtraction will help in the construction of minimal, stable, 
and efficient subtraction algorithms, taking maximal advantage of existing 
analytic information.}

\FullConference{13th International Symposium on Radiative Corrections\\
		24-29 September, 2017\\
		St. Gilgen, Austria}

\begin{document}


\section{Introduction}
\label{intro}

The infrared structure of virtual corrections to perturbative gauge theory 
amplitudes is very well understood~\cite{Catani:1998bh,Sterman:2002qn,
Dixon:2008gr,Gardi:2009qi,Gardi:2009zv,Becher:2009cu,Becher:2009qa}: 
soft and collinear singularities associated with on-shell configurations of 
loop momenta are known to factorise from the hard scattering, and they 
follow a pattern of exponentiation dictated by a small set of soft and collinear 
anomalous dimensions, which (in the massless case) are known to three 
loops~\cite{Almelid:2015jia}. Similarly, the factorisation properties of real 
radiation amplitudes, in the limits when one or more external partons become 
soft, or collinear to each other, are understood in considerable generality
\cite{Kosower:1999xi,Catani:1998nv,Catani:1999ss}, and all the splitting 
kernels (at the amplitude level) necessary for next-to-next-to-leading 
order (NNLO) calculations have been computed~\cite{Catani:1999ss,
Campbell:1997hg,Bern:1999ry,Catani:2000pi}.

Notwithstanding these early results, the problem of constructing general 
and efficient algorithms for the calculation of infrared-safe observables 
in hadronic scattering processes beyond NLO has proven to be very 
challenging. After pioneering results on jet production in electron-positron 
annihilation~\cite{GehrmannDeRidder:2008ug,Weinzierl:2008iv}, several 
approaches have been developed to tackle the problem for hadron scattering; 
some have already been successfully deployed to selected processes at 
NNLO, while some are in different stages of development: several talks 
at this conference have discussed recent developments of the various 
proposed methods and their applications~\cite{Cacciari,Torrielli,Roentsch,
Behring,Kallweit,Kardos,Rodrigo,Gehrmann}. To give an extremely concise 
summary of the state of the art, we note that the first fully hadronic process 
to be studied at NNLO at the level of fully differential distributions was the 
production of top-antitop pairs~\cite{Czakon:2015owf,Czakon:2016ckf}, 
a seminal result achieved using a predominantly numerical subtraction 
scheme; processes with one final-state jet, or with electroweak final states, 
have subsequently been tamed using phase-space slicing methods such 
as N-Jettiness subtraction~\cite{Boughezal:2015dra,Boughezal:2015ded,
Boughezal:2016wmq}, or $q_T$ subtraction~\cite{Grazzini:2017mhc,
Grazzini:2017ckn,deFlorian:2016uhr}; very recently, the first complete 
results for dijet production (in the planar limit) have been presented
\cite{Currie:2017eqf}, using Antenna subtraction. Other methods have 
been applied using process-specific approximations~\cite{Cacciari:2015jma}, 
or to processes with electroweak initial states~\cite{Tulipant:2017ybb,
DelDuca:2016csb,DelDuca:2016ily}, while yet other methods are at the 
design stage~\cite{Sborlini:2016hat}, or undergoing preliminary testing
\cite{Caola:2017dug}. Finally, it is worth pointing out that limited extensions 
to N$^3$LO ~\cite{Dulat:2017prg} are being attempted.

All the methods devised and implemented so far are very demanding, 
either in terms of the necessary numerical calculations, or because of 
the intricacy of the analytical integrations to be performed, or both, underscoring 
the difficulty of the problem. At the same time, the massive amounts of data 
being delivered by LHC, and the increasing requirements on the precision 
of theoretical predictions for a widening array of processes, point to the 
necessity of preparing for much more intricate applications. In the coming 
years, differential calculations for multi-jet final states at NNLO, and for 
selected observables at N$^3$LO, are likely to become both feasible 
and relevant. In light of current results, this kind of generalisation will 
likely require a massive optimisation and streamlining of current subtraction 
methods, and possibly the development of entirely new ones.

In the present contribution, we would like to present some preliminary 
results in this direction. Specifically, we explore the relationship between 
the factorisation of IR divergences in virtual corrections to scattering 
amplitudes, and the local counterterms required for infrared subtraction. 
This investigation has a two-fold purpose. On the one hand, as we will 
see, it provides a `shopping list' of gauge-invariant operator matrix 
elements which yield soft and collinear counterterms to any order in 
perturbation theory; in principle, this allows to decouple the identification 
of the counterterms from the calculation of specific scattering amplitudes, 
in contrast to the methods followed so far at the two-loop level (see, for 
example,~\cite{Duhr:2013msa,Li:2013lsa}). On the other hand, by tracing 
an explicit connection between virtual corrections and real radiation, one 
may hope that the remarkable simplifications inherent in virtual factorisation 
may be reflected in the properties of the counterterms: for example, high 
orders in the virtual soft and jet functions are tied to low orders by exponentiation,
and furthermore it is known a priori that entire classes of contributions to 
soft anomalous dimensions vanish to all orders in perturbation theory
\cite{Gardi:2009qi,Becher:2009qa}. We will begin, in \secn{facto}, by 
very briefly reviewing what is known about the IR factorisation of fixed-angle 
multi-parton scattering amplitudes; we will then present a basic outline of 
the subtraction problem in \secn{outli}, focusing for simplicity on the NLO 
case; in Sections~\ref{soft} and \ref{colli}, we present and motivate our 
general ansatz for soft and collinear subtraction counterterms to all orders; 
in \secn{NLOs}, as an example, we show how our definitions combine to 
reconstruct the well-understood structure of final-state IR subtraction at 
NLO; finally, we give a brief perspective in \secn{persp}.


\section{Infrared structure of virtual corrections}
\label{facto}

Massless multi-particle gauge theory scattering amplitudes are affected
by soft and collinear divergences beyond leading order in perturbation 
theory. Divergences arise from long-distance contributions to virtual 
corrections, and therefore enjoy a degree of universality, being largely 
independent from the details of the hard scattering. Divergent contributions 
can therefore be expressed in terms of universal, gauge-invariant operator 
matrix elements, multiplying a finite hard remainder. The final result can
be derived using diagrammatic techniques~\cite{Sterman:1995fz}, or
effective field theory methods~\cite{Feige:2014wja}, and can be expressed
in terms of an all-order factorisation theorem. Writing the $n$-particle scattering 
amplitude as $A^{a_1 \ldots a_n}$, where the colour indices $a_i$ can 
belong to any representation of the gauge group, the first step is to choose
a basis in the space of available color structures for the given process, by 
picking a suitable set of color tensors $c_K^{a_1 \ldots a_n}$. The scattering
amplitude can then be expressed in terms of its components $A_K$ in this 
basis. These `partial amplitudes' obey the factorisation
\begin{eqnarray}
  A_{K} \left(p_i/\mu \right) \, = \, 
  \prod_{i = 1}^n \left[ {\displaystyle J_i 
  \left( \frac{(p_i \cdot n_i)^2}{n_i^2 \mu^2} \right)} \!\!
  \Bigg/ \!\!\! {\displaystyle {\cal J}_i \left( \frac{(\beta_i \cdot n_i)^2}{n_i^2}\right)} 
  \, \right] {\cal S}_K^{\,\,\, L} \left( \beta_i \cdot \beta_j \right) \,  
  H_{L} \left( \frac{p_i \cdot p_j}{\mu^2},
  \frac{(p_i \cdot n_i)^2}{n_i^2 \mu^2}\right) ,
\label{factor}
\end{eqnarray}
where for simplicity we have suppressed the dependence on the renormalised 
coupling $\alpha_s(\mu^2)$ and on the dimensional regularisation parameter 
$\epsilon$, and where $p_i$, $i = 1, \ldots, n$, are the external momenta, while
$\beta_i$ are the corresponding dimensionless four-velocities, obtained by 
rescaling the momenta by a common hard scale, for example setting $p_i = 
\mu \beta_i$.

The `jet' functions $J_i$ in \eq{factor} collect all collinear singularities associated 
with virtual quanta emitted in the direction of particle $i$; they depend on the 
particle spin, but they are `colour singlet' quantities, in the sense that they 
do not affect the color of the external particle; they can be defined as operator 
matrix elements involving Wilson lines. For example, in the case of an outgoing 
quark, we can use
\beq
  \overline{u}(p) \, J \left( \frac{(p \cdot n)^2}{n^2 \mu^2} \right) \, = \, 
  \langle p \, | \, \overline{\psi} (0) \, \Phi_n (0, \infty) \,  | 0 \rangle \, ,
\label{Jdef}
\eeq
where $\Phi_n$ is a semi-infinite Wilson line in an arbitrary direction $n_i^\mu$,
with $n^2 \neq 0$, according to the definition
\beq
  \Phi_n (\lambda_2, \lambda_1) \, \equiv \, {\cal P} \exp \left[ {\rm i} g 
  \int_{\lambda_1}^{\lambda_2} d \lambda \, n \cdot A (\lambda n) \right] \, .
\label{Wline}
\eeq
The vectors $n_i$ play the role of `factorisation vectors', and they ensure 
gauge invariance of the operator matrix element defining $J_i$. The
specific functional dependence of $J_i$ on its argument is dictated by
the masslessness of the external particle, $p_i^2 = 0$, and by the 
rescaling invariance of the semi-infinite Wilson line under $n_i^\mu \to \,
\kappa_i \, n_i^\mu$, with $\kappa_i$ a constant.

The colour-singlet nature of the jet functions $J_i$ follows from collinear 
power counting at the level of Feynman graphs, and greatly simplifies
the structure of the factorisation. Soft singularities, on the other hand, 
are not color diagonal, and therefore they are organised in a matrix, 
${\cal S}_{L K}$, which is purely eikonal. Indeed, since soft gluons have
long wavelength, they do not resolve the details of the hard interaction 
nor the internal structure of the jets, and therefore they couple effectively 
to Wilson lines in the colour representations of the corresponding hard 
external partons, and oriented along their four-velocities $\beta_i$. In 
a chosen basis of independent tensors $c_K$ in colour space, one defines
\beq
  \left( c_L \right)^{ \{ a_k \} } {\cal S}^L_{\,\,\,\, K} \left(\beta_i 
  \cdot \beta_j \right) \, = \, \langle 0 |  \, \prod_{k = 1}^n \Big[ 
  \Phi_{\beta_k} (\infty, 0)^{ \,\, a_k \, b_k } \Big]  \, | 0 \rangle \, 
  \left( c_K \right)_{ \{b_k \} } \, .
\label{softcorr}
\eeq
Notice that the dependence of the soft matrix on the four-velocities 
$\beta_i$ is severely constrained by the rescaling invariance of the
semi-infinite Wilson line operators under $\beta_i \to \kappa_i \beta_i$:
in the massless case, only dependence on conformal invariant cross 
ratios of the form
\beq
  \rho_{i j k l } \, \equiv \, \frac{\beta_i \cdot \beta_j \, \beta_k \cdot 
  \beta_l}{\beta_i \cdot \beta_l \, \beta_j \cdot \beta_k}
\label{ccrs}
\eeq
would be allowed, except for the fact that the rescaling invariance
of the correlator in \eq{softcorr} is broken by collinear poles, leading
to an explicit but highly constrained dependence on $\beta_i \cdot 
\beta_j$, proportional to the cusp anomalous dimension~\cite{Gardi:2009qi,
Gardi:2009zv,Becher:2009cu,Becher:2009qa}.

The final ingredients of the factorisation formula \eq{factor} are the soft 
approximations of the jet functions $J_i$, sometimes called call `eikonal 
jets'. They are defined by
\beq
  {\cal J}_i \left( \frac{(\beta_i \cdot n_i)^2}{n_i^2} \right)  \, = \, 
  \langle 0 | \, \Phi_{\beta_i}(\infty, 0) \, 
  \Phi_{n_i} (0, \infty) \, | 0 \rangle~.
\label{calJdef}
\eeq
Introducing eikonal jets is necessary in order to avoid double counting
of gluons that are both soft and collinear to one of the hard external partons:
such gluons appear both in the jet functions $J_i$ and in the soft matrix 
${\cal S}$. It is however simple to subtract this double counting: one just 
needs to divide each jet $J_i$ by its own soft approximation ${\cal J}_i$,
as done in \eq{factor}. Notice once again the argument of the eikonal jet
functions ${\cal J}_i$: we allow only for homogeneous dependence on 
the (non-light-like) vectors $n_i$, while non-homogeneous dependence
on the light-like vectors $\beta_i$ is allowed, as was the case for the 
soft matrix ${\cal S}$.

The content of the factorisation theorem in \eq{factor} is that all soft and 
collinear divergences in the amplitude, to all orders in perturbation theory, 
are generated by the universal soft and jet functions: the vector of hard 
functions $H_K$ is then a matching coefficient, collecting all finite remainders, 
and it is finite as $\e \to 0$. The theorem is powerful because, like all factorisations,
it implies the existence of evolution equations, which can be solved to
construct a resummation (exponentiation) of infrared poles. To briefly 
illustrate this fact, we note that \eq{factor} can be rewritten by collecting 
soft and collinear factors in a form reminiscent of ultraviolet renormalisation,
as~\cite{Gardi:2009zv,Becher:2009cu}
\beq 
  {\cal A} \left( \frac{p_i}{\mu} \right) \, = \, {\cal Z} \left( \frac{p_i}{\mu} \right)
  {\cal H} \left( \frac{p_i}{\mu} \right) \, ,
\label{IRfact}
\eeq
where the cancellation of the dependence on the factorisation vectors $n_i$ 
between the various factors comprising ${\cal Z}$ and ${\cal H}$ has already 
been implemented, the scattering amplitude ${\cal A}$ is a vector in color 
space, and ${\cal Z}$ is a matrix encoding all infrared divergences, acting 
upon the finite vector ${\cal H}$. The infrared operator ${\cal Z}$ obeys a 
renormalisation group equation which, in dimensional regularization
\cite{Magnea:1990zb} and for $d = 4 - 2 \epsilon > 4$, has a very simple 
solution, expressed in terms of a soft anomalous dimension matrix $\Gamma$ 
as
\beq
  {\cal Z} \left(\frac{p_i}{\mu}, \alpha_s (\mu^2) \right) \, = \,  
  {\cal P} \exp \left[ \frac{1}{2} \int_0^{\mu^2} \frac{d \lambda^2}{\lambda^2} \, \,
  \Gamma \left(\frac{p_i}{\lambda}, \alpha_s(\lambda^2) \right) \right] \, .
\label{RGsol}
\eeq
It is clear from this discussion that the matrix $\Gamma$ is the key ingredient 
for the solution of the perturbative IR problem: it is currently known at three loops
for massless particles~\cite{Almelid:2015jia}. In the context of the subtraction 
problem, we note that the exponentiation in \eq{RGsol}, and the highly non-trivial
form of the matrix $\Gamma$~\cite{Gardi:2009qi,Becher:2009qa,Almelid:2015jia},
imply a number of connections between virtual IR poles at different orders, as
well as significant constraints on their structure. These connections and constraints
must at some level be reflected by the real radiation counterterms: elucidating
this relationship is one of the goals of our approach.


\section{Infrared subtraction: an outline}
\label{outli}

In order to begin our discussion, let us start with a very concise summary of 
the subtraction procedure at lowest non-trivial order. For the sake of simplicity,
we will focus on final state radiation only. We consider a generic infrared-safe
observable $O$, which receives its leading order contribution from a final
state with $n$ partons. For such an observable, NLO distributions are in
principle computed as
\beq
  \langle O \rangle_{\NLO} \, = \, \lim_{d \to 4} 
  \left\{ \int d \pn^\dd \big[ B_n^\dd + V_n^\dd \big] O_n
  + \int d \pnpo^\dd R_\npo^\dd \, O_\npo \right\} \, ,
\label{eq:ONLO}
\eeq
where $B_n$ is the Born-level squared amplitude, $V_n$ the corresponding 
one-loop corrections, and $R_\npo$ is the single-real-radiation tree-level 
squared amplitude. The observable $O$ admits explicit expressions $O_m$ 
in the $m$-particle phase space, and IR safety requires that $O_{n+1} \to 
O_n$ in all degenerate limits where one particle becomes unresolved, becoming 
soft or collinear to another one. The practical problem in evaluating \eq{eq:ONLO}
is that the {\it  r.h.s.} of \eq{eq:ONLO} must in general be evaluated numerically
(due to the complexity of the typical observable $O$ and of the relevant matrix 
elements), while the explicit poles in the first term must be cancelled against 
singularities arising in the phase-space integration of the second term. The 
idea of subtraction is to introduce a set of {\it local counterterm} functions 
$K_m$, with the property that they reproduce the singular behaviour of the 
real-radiation squared matrix element $R_{n+1}$ in all unresolved limits, 
while at the same time being simple enough to be analytically integrated 
over the single-unresolved-particle phase space. One then defines
\beq
  \langle O \rangle_{\rm ct} \, = \, \int d \pn \, d \poh^\dd \, K_\npo^\dd \, O_n \, ,
\label{eq:C}
\eeq
and then proceeds to compute analytically the integral
\beq
  I^\dd_n \, = \, \int d \poh^\dd \, K^\dd_\npo \, .
\label{intctnlo}
\eeq
\eq{eq:ONLO} can now be rewritten identically as 
\beq
  \hspace{-2mm}
  \langle O \rangle_{\NLO} = \int d \pn \left[ B_n^\fd + 
  \left( V_n + I_n \right)^\fd \right] O_n + \int d \pn \left[\int \!
  d \po^\fd R_\npo^\fd O_\npo - \int \! d \poh^\fd K_\npo^\fd O_n
  \right] .
\label{eq:subtONLO}
\eeq
By the standard theorems concerning the cancellation of IR singularities, the
two terms in \eq{eq:subtONLO} are now separately finite and can be evaluated
numerically in $d = 4$. 

Clearly, at higher orders this relatively straightforward procedure becomes 
much more intricate, due to the presence of a large number of overlapping
singular regions. Furthermore, this sketchy formal treatment neglects the
technical difficulties associated with the need to construct precise and efficient
phase space mappings relating the radiative and the Born configurations: our 
viewpoint on these issues is discussed in~\cite{Torrielli}. In what follows, 
we focus on the general structure of the counterterm functions $K_m$.
As we will see below, the fact that the structure of virtual poles is known 
to all orders in terms of gauge invariant operator matrix elements suggests
a completely general definition of soft and collinear counterterms to any 
order; furthermore, the structure of the virtual factorisation also dictates 
a pattern of cancellation of overlapping singularities, in principle allowing
for an automatic construction of the subtraction procedure at NNLO and 
beyond.


\section{Local soft counterterms}
\label{soft}

Let us begin by focusing on purely soft singular configurations. Real soft 
radiation at leading power is described by replacing hard radiating 
particles with Wilson lines, so one is naturally led to consider the matrix 
elements for the radiation of $m$ soft partons from $n$ (outgoing) Wilson 
lines,
\beq
  {\cal S}_{\lambda_1 \ldots \lambda_m} \left(k_1, \ldots, k_m; \beta_i \right)
  & \equiv & \bra{k_1, \lambda_1; \ldots; k_m, \lambda_m} \, \prod_{i = 1}^n 
  \Phi_{\beta_i} (\infty, 0) \, \ket{0} \nonumber \\
  & \equiv & g^m \, \epsilon^{* \mu_1}_{\lambda_1} (k_1) \ldots 
  \epsilon^{* \mu_m}_{\lambda_m} (k_m) \, J^{\cal S}_{\mu_1 \ldots \mu_m}
  \left(k_1, \ldots, k_m; \beta_i \right) \nonumber \\
  & \equiv &  g^m \, \sum_{p = 0}^\infty \, \left( \frac{\alpha_s}{\pi} \right)^p \,
  {\cal S}_{\lambda_1 \ldots \lambda_m}^{(p)} \left(k_1, \ldots, k_m; \beta_i \right) \, ,
\label{softrad}
\eeq
where in the second line we have extracted polarisation vectors\footnote{Here 
we consider only radiated gluons, but it is straightforward to generalise the 
definition to quarks.} and an overall power of the coupling, in order to define
a multi-gluon soft emission current $J^{\cal S}_{\mu_1 \ldots \mu_m}$, while 
in the third line we have defined the perturbative coefficients of the loop 
expansion of our matrix element.

We note that existing finite-order calculations and all-order arguments are
consistent with the soft factorisation of the radiative matrix element
\beq
  {\cal A}_{\lambda_1 \ldots \lambda_m} \left(k_1, \ldots, k_m; p_i \right)
  \, \simeq \, {\cal S}_{\lambda_1 \ldots \lambda_m} \left(k_1, \ldots, k_m; 
  \beta_i \right) \, {\cal H}(p_i) \, ,
\label{ampfact}
\eeq
with corrections that are finite in dimensional regularisation and integrable 
in the soft-gluon phase space. Squaring the matrix element and (optionally) 
summing over polarisations one can then write 
\beq
  \sum_{\lambda_1 \ldots \lambda_m} \left| {\cal A}_{\lambda_1 
  \ldots \lambda_m} \left(k_1, \ldots, k_m; p_i \right) \right|^2 \, \simeq \, 
  S_m \left(k_1, \ldots, k_m; \beta_i \right) \, \left| {\cal H}(p_i) \right|^2 \, ,
\label{squampfact}
\eeq
which defines the cross-section-level radiative soft function
\beq
  S_m \left(k_1, \ldots, k_m; \beta_i \right) & \equiv & 
  \sum_{\lambda_1 \ldots \lambda_m} 
  \bra{0} \prod_{i = 1}^n 
  \Phi_{\beta_i} (0, \infty) \ket{k_1, \lambda_1; \ldots; k_m, \lambda_m}
  \bra{k_1, \lambda_1; \ldots; k_m, \lambda_m} \prod_{i = 1}^n 
  \Phi_{\beta_i} (\infty, 0) \ket{0}  \nonumber \\
  & \equiv &
  \left( 4 \pi \alpha_s \right)^m \sum_{p = 0}^\infty \left( \frac{\alpha_s}{\pi} \right)^p 
  S_m^{(p)} \left(k_1, \ldots, k_m; \beta_i \right) \, .
\label{softsigma}
\eeq
It is clear that the perturbative coefficients $S_m^{(p)} \left(k_1, \ldots, k_m; 
\beta_i \right)$ are ideally suited to act as minimal local counterterms for the 
soft singular regions of phase space, to any order in perturbation theory. 
Indeed, summing over the number of particles in the final state and integrating
over their phase space, by completeness we find
\beq
  \sum_{m = 0}^\infty \int d \Phi_m (k_j) \, S_m \left(k_1, \ldots, k_m; 
  \beta_i \right) \, = \, 
  \bra{0} \prod_{i = 1}^n 
  \Phi_{\beta_i} (0, \infty) \prod_{i = 1}^n 
  \Phi_{\beta_i} (\infty, 0) \ket{0} \, .
\label{complete}
\eeq
The right-hand side is effectively a total eikonal cross section, and thus finite 
order by order by the general theorems concerning the cancellation of IR
divergences. The $m = 0$ term gives the purely virtual contribution, and 
one easily verifies diagrammatically that real radiation corrections construct
order by order the familiar pattern of cancellations.

In order to illustrate the results, and to make contact with earlier work, in 
particular~\cite{Catani:1999ss,Catani:2000pi}, consider first the tree-level
radiative amplitude with a single soft gluon. \eq{ampfact} in this case 
reduces to
\beq
  {\cal A}^{(0)}_\lambda (k, p_i) \, = \, g \, \epsilon^*_{\lambda} (k) \cdot 
  J^{\cal S}_{(0)} (k, \beta_i) \, {\cal H}^{(0)} (p_i) + {\cal O}(k^0) \, .
\label{treesoft}
\eeq
The definition in \eq{softrad} at this level reduces to
\beq
  g \, \epsilon^*_{\lambda} (k) \cdot J^{\cal S}_{(0)} (k, \beta_i) \, = \, 
  \left. \bra{k, \lambda} \, \prod_{i = 1}^n \Phi_{\beta_i} (\infty, 0) \, 
  \ket{0} \right|_{\rm tree \, level} \, ,
\label{treesoftcurr}
\eeq
which immediately yields the well-known result
\beq 
  J^{\cal S, \, \mu}_{(0)} (k, \beta_i) \, = \, \sum_{i = 1}^n \,
  \frac{\beta_i^\mu}{\beta_i \cdot k} \, {\bf T}_i \, \, .
\label{treesoftcurexp}
\eeq
Squaring the tree amplitude in \eq{treesoft} and summing over soft-gluon 
polarisations yields the factorisation of the standard eikonal prefactor
\beq
  \sum_\lambda \left| {\cal A}^{(0)}_\lambda (k, p_i) \right|^2 & \simeq & 
  - \, 4 \pi \alpha_s \, \sum_{i,j = 1}^n \, \frac{\beta_i \cdot \beta_j}{\beta_i \cdot k \, 
   \beta_j \cdot k} \, {\cal A}^{(0) \dagger} (p_i) \, {\bf T}_i \cdot {\bf T}_j \, 
  {\cal A}^{(0)} (p_i) \nonumber \\ & = & 
  S_1^{(0)} \left(k; \beta_i \right) \, \left| {\cal H}^{(0)} (p_i) \right|^2 \, .
\label{treesqu}
\eeq
In the second line of \eq{treesqu} we have adopted a schematic operator 
notation, defined by the first line: the one-gluon tree-level radiative soft function
is a color operator acting on the colour-correlated Born squared matrix element,
which at tree level is finite and coincides with the hard part. As one might expect,
the two-gluon soft current can similarly be computed, recovering the results
of~\cite{Catani:1999ss}, and multiple soft gluon currents follow the same pattern.

The picture gets a bit more interesting at one loop. Considering as an example
single-gluon emission, one can compare the factorisation in \eq{ampfact} with
the one proposed in Ref.~\cite{Catani:2000pi}. They read
\beq
  {\cal A}_\lambda \left(k ; p_i \right) \, \simeq \, {\cal S}_\lambda
  \left(k; \beta_i \right) \, {\cal H}(p_i) \, ,
  \qquad \qquad
  {\cal A}_\lambda \left(k; p_i \right) \, \simeq \, g \, \epsilon^*_{\lambda} (k) \cdot 
  J_{\scriptscriptstyle{CG}} \left( k, \beta_i \right) \, {\cal A} (p_i) \, ,
\label{twofact}
\eeq
respectively, where we denoted with $J_{\scriptscriptstyle{CG}}$ the 
Catani-Grazzini soft-gluon current. Intuitively, the difference is that virtual 
IR divergences are still contained in the non-radiative matrix element ${\cal A}$ 
in the Catani-Grazzini factorisation, whereas the hard factor ${\cal H}$ is finite. 
It is however easy to map the two expressions in \eq{twofact}: expanding the 
one-loop amplitude in the first expression in \eq{twofact} we find 
\beq
  {\cal A}_\lambda^{(1)} \left(k ; p_i \right) \, \simeq \, {\cal S}_\lambda^{(0)}
  \left(k; \beta_i \right) \, {\cal H}^{(1)} (p_i) \, + \,  
  {\cal S}_\lambda^{(1)} \left(k; \beta_i \right) \, {\cal H}^{(0)} (p_i) \, .
\label{expone}
\eeq
We can then express the one loop non-radiative hard part in terms of the full
amplitude using \eq{ampfact} for $m = 0$, obtaining
\beq
  {\cal A} \left( p_i \right) \, \simeq \, {\cal S} \left( \beta_i \right) \, {\cal H} (p_i)  
  \quad \longrightarrow \quad
  {\cal H}^{(1)} (p_i) \, = \, {\cal A}^{(1)} \left( p_i \right) -
  {\cal S}^{(1)} \left( \beta_i \right) \, {\cal A}^{(0)} (p_i) \, ,
\label{exptwo}
\eeq
where for simplicity we reabsorbed the tree-level soft function, a pure color 
factor, into ${\cal H}^{(1)}$. Substituting this result in \eq{twofact}, and solving 
for the one-loop soft current, we find
\beq
  g^3 \, \epsilon^*_\lambda (k)  \cdot J^{(1)}_{\scriptscriptstyle{CG}} (k, \beta_i) \, 
  {\cal A}^{(0)} (p_i) \, = \, \left[ {\cal S}_\lambda^{(1)} \left(k; \beta_i \right) - 
  {\cal S}_\lambda^{(0)} \left(k; \beta_i \right) {\cal S}^{(1)} \left( \beta_i \right)
  \right] \, {\cal A}^{(0)} (p_i) \, .
\label{CGosc}
\eeq
Working out the diagrammatics, it is not difficult to recover the full result of 
Ref.~\cite{Catani:2000pi} for the one-loop current: in a sense, \eq{CGosc}
provides a transparent interpretation of the cancellations that lead to the
purely non-abelian structure of $J^{(1)}_{\scriptscriptstyle{CG}}$ first derived
in~\cite{Catani:2000pi}. It is not difficult to extend these structural arguments
to either more gluons ore more loops, obtaining precise operator definitions
of the corresponding soft currents: this can lead, for example, to a first-principle 
calculation of the two-loop soft-gluon current first computed in~\cite{Duhr:2013msa,
Li:2013lsa}, and to a generalisation of that result beyond the two-hard-parton case.


\section{Local collinear counterterms}
\label{colli}

Local collinear counterterms can be constructed with the same method, 
starting from the definition of virtual jets, and suitably allowing for final 
state radiation. More precisely, for example in the case of a final state 
quark jet, consider the matrix elements\footnote{The amplitude-level 
radiative jets defined in \eq{qradjet} are closely related (but not identical) 
to the ones introduced in Refs.~\cite{DelDuca:1990gz,Bonocore:2015esa,
Bonocore:2016awd}.}
\beq
  {\cal J}_{m, \, s}^{\lambda_j} \left(k_j; p, n \right)
  & \equiv & \bra{p, s; k_j, \lambda_j}  \overline{\psi} (0) \, 
  \Phi_{n} (0, \infty) \ket{0} \, \equiv \, g^m \sum_{p = 0}^\infty 
  \left( \frac{\alpha_s}{\pi} \right)^p {\cal J}_{m, s, \lambda_j}^{(p)} 
  \left(k_j; p, n \right) \, ,
\label{qradjet}
\eeq
where $s$ is the quark spin and $\lambda_j$ are the gluon polarisations. 
Note that, at this stage, spinors and polarisation vectors for final-state partons
are included in the definitions of ${\cal J}$ and of its perturbative coefficients 
${\cal J}^{(p)}$: one could extract them, as done in the second line of 
\eq{softrad}, and define `collinear currents' for the emission of a given 
number of collinear final-state partons. Because of the non-trivial momentum 
flow in the collinear limit, rather than simply squaring the radiative jet amplitude 
in \eq{qradjet} it is appropriate to introduce a shift in the complex conjugate 
amplitude and then perform a Fourier transform, defining cross-section-level 
radiative jets as
\beq
  J_m^{s, \lambda_j} \left(k_j; l, p, n \right) & \equiv & 
  \int d^d x \, {\rm e}^{{\rm i} l \cdot x} \, \bra{0} 
  \Phi_n (\infty, x) \, \psi(x) \ket{p, s; k_j, \lambda_j}
  \bra{p, s; k_j, \lambda_j} \overline{\psi} (0) \, 
  \Phi_n (0, \infty) \ket{0}  \nonumber \\
  & \equiv &
  \left( 4 \pi \alpha_s \right)^m \sum_{p = 0}^\infty \left( \frac{\alpha_s}{\pi} \right)^p 
  J_{m, s, \lambda_j}^{(p)} \left(k_j; l, p, n \right) \, .
\label{qradjetsq}
\eeq
The perturbative coefficients of the radiative jets, defined in the second line 
of \eq{qradjetsq}, are natural possible definitions of collinear counterterms. 
Indeed, applying again completeness, one finds that
\beq
  & & \sum_{m = 0}^\infty \int d \Phi_{m+1} \left(p, k_j \right) \, 
  \sum_{\{ \lambda_j\}} J_{m, s, \lambda_j} \left(k_j; l, p, n \right) 
  \, = \, \nonumber \\
  & & \hspace{2cm} = \, {\rm Disc} \left[ \int d^d x \, {\rm e}^{{\rm i} l \cdot x} \,
  \bra{0} \Phi_n (\infty, x) \psi(x) \overline{\psi} (0) \Phi_n (0, \infty) \ket{0} \right] \, .
\label{compcoll}
\eeq
The right-hand side of \eq{compcoll} is a generalised two-point function (closely 
related to the inclusive jet function defined in~\cite{Sterman:1986aj}) and it is 
manifestly finite, since it is fully inclusive in the final state. Cross-section-level
eikonal jets can be defined analogously, as
\beq
  J_{m, \lambda_j}^{\rm E} \left(k_j; l, \beta, n \right) & \equiv & 
  \int d^d x \, {\rm e}^{{\rm i} l \cdot x} \, \bra{0}  
  \Phi_n (\infty, x) \Phi_\beta (x, \infty) \ket{k_j, \lambda_j}
  \bra{k_j, \lambda_j} \Phi_\beta (\infty, 0)
  \Phi_n (0, \infty) \ket{0} \nonumber \\
  & \equiv &
  \left( 4 \pi \alpha_s \right)^m \sum_{p = 0}^\infty \left( \frac{\alpha_s}{\pi} \right)^p 
  J_{m, \lambda_j}^{{\rm E}, \, (p)} \left(k_j; l, \beta, n \right) \, ,
\label{eikjetsq}
\eeq
where now the Fourier transform simply sets the total final state momentum of
soft-collinear gluons to be $l^\mu$, and once again polarisation vectors have 
been included in the definition. Note that eikonal jets, as expected, do not 
depend on the spin of the hard emitting parton.

As a sanity check, one can compute the single-gluon radiative jet: at 
cross-section level, summing over polarisations, one should recover the 
(unpolarised) Altarelli-Parisi splitting function $P_{qq}$. Indeed, a straightforward 
calculation yields
\beq
  \sum_{s, \lambda} J_1^{s, \lambda} \left( k; l, p, n \right) \, = \, 
  \frac{4 \pi \alpha_s C_F}{l^2} \, (2 \pi)^d \, \delta^d \left(l - p - k \right) \,
  \left[ - \slash{l} \gamma_\mu \slash{p} \gamma^\mu \slash{l} \, + \,
  \frac{1}{k \cdot  n} \left( \slash{l} \slash{n} \slash{p} + \slash{p} \slash{n} 
  \slash{l} \right) \right] \, ,
\label{onecollglu}
\eeq
up to terms proportional to $n^2$. Introducing a Sudakov parametrisation
\beq
  p^\mu \, = \, z l^\mu + {\cal O} \left( l_\perp \right) \, , \qquad
  k^\mu  \, = \, (1 - z) l^\mu + {\cal O} \left( l_\perp \right) \, , \qquad
  n^2 \, = \, 0 \, ,
\label{Sudakov}
\eeq
and taking the collinear limit $l_\perp \to 0$ one easily finds
\beq
  \sum_{s, \lambda} J_1^{s, \lambda} \left( k; l, p, n \right) \, = \, 
  \frac{8 \pi \alpha_s C_F}{l^2} \, (2 \pi)^d \, \delta^d \left(l - p - k \right) \,
  \left[ \frac{1 + z^2}{1 - z} - \epsilon \left( 1 - z \right) \, + \, 
  {\cal O} \left( l_\perp \right) \right] \, .
\label{AP0}
\eeq
Once again, a detailed mapping to the axial-gauge calculation of 
Ref.~\cite{Catani:1999ss} is possible: diagrams involving gluon emission 
from the Wilson lines in our calculation reconstruct the non-trivial
terms arising from the axial-gauge gluon propagator in~\cite{Catani:1999ss}.

We note that, unlike the situation for soft radiation, the collinear counterterms
defined by \eq{qradjetsq} are not minimal: the total final state momentum $l^\mu$
is generic, and in principle the collinear limit must be taken at the end of the 
calculation. Furthermore, while the matrix elements are gauge-invariant, they 
are $n$-dependent: at loop level, choosing $n^2 = 0$ from the outset would 
need to be done with care, since in this case there are spurious collinear 
divergences in loops associated with emissions from the Wilson line (see
Ref.~\cite{Bonocore:2015esa} for a discussion of this issue). In short, there
is room for improvement in our definition of collinear counterterms, and a 
refinement is likely to be practically necessary when tackling high-order 
corrections.


\section{NLO subtraction: a sketch}
\label{NLOs}

We conclude this brief discussion with an outline of how the framework we have 
introduced translates into a sequence of steps to construct a subtraction scheme.
Here we will only describe very schematically what happens at NLO, a more
detailed account will be found in~\cite{Us}. The main idea, as discussed in the 
Introduction, is to start from the structure of IR divergences in the virtual contribution
to the observable distribution. We consider then the non-radiative scattering 
amplitude at the one-loop level, which is given by
\beq
  {\cal A} (p_i) & = &  {\cal S}^{(0)} (\beta_i) {\cal H}^{(0)} (p_i) \, + \,
  \frac{\alpha_s}{\pi} \bigg[
  {\cal S}^{(1)} (\beta_i) {\cal H}^{(0)} (p_i) \, + \,
  {\cal S}^{(0)} (\beta_i) {\cal H}^{(1)} (p_i) \nonumber \\ && \qquad \quad + \, 
  \sum_i \left( {\cal J}^{(1)} (p_i) - {\cal J}_E^{(1)} (\beta_i) \right)  
  {\cal S}^{(0)} (\beta_i) {\cal H}^{(0)} (p_i) \bigg] \, + \, {\cal O} \left( \alpha_s^2 \right) \, .
\label{oneloopamp}
\eeq 
The virtual term in \eq{eq:ONLO} is then
\beq
  V_n \, \equiv \, 2 \, {\bf Re} \Big[ {\cal A}_n^{(0) *} {\cal A}_n^{(1)} \Big] & = & 
  \frac{\alpha_s}{\pi} \, S_0^{(1)} (\beta_i) \left| {\cal H}^{(0)}_n (p_i) \right|^2 
  \nonumber \\ &&   \quad + \, 
  \frac{\alpha_s}{\pi} \, \sum_i \left( J_0^{(1)} (p_i) - J_{E, 0}^{(1)} (\beta_i) \right)
  \left| {\cal H}_n^{(0)} (p_i) \right|^2 \, + \, {\rm finite} \, .
\label{virtol}
\eeq
To identify the real-radiation counterterms we now just need to expand the
completeness relations for soft and collinear functions to NLO. At this order,
they imply simply that
\beq
  S_0^{(1)} (\beta_i) \, + \, 4 \pi^2 \int d \Phi_2^{(d)} \, S_1^{(0)} (k, \beta_i)
  \, = \, {\rm finite} 
\label{NLOsoftfin}
\eeq
and
\beq
  J_0^{(1)} (l, p, n) \, + \, 4 \pi^2 \int d \Phi_2^{(d)} \, J_1^{(0)} (k; l, p, n)
  \, = \, {\rm finite} \, ,
\label{NLOcollfin}
\eeq
where we have summed over polarisations for simplicity. Matching to 
\eq{oneloopamp}, this means that local soft and collinear counterterms 
can be constructed according to
\beq
  K_{n+1}^{\rm soft} \, = \, 4 \pi \alpha_s \, S_1^{(0)} (k, \beta_i) \, 
  \left| {\cal H}^{(0)}_n (p_i) \right|^2 \, ,
\label{NLOsoftct}
\eeq
while
\beq
  K_{n+1}^{\rm coll.} \, = \, 4 \pi \alpha_s \, \sum_i J_1^{(0)} (k_i; l, p_i, n_i) \, 
  \left| {\cal M}^{(0)}_n \left( p_1, \ldots, p_{i - 1}, l, p_{i +1}, \ldots, p_n \right) 
  \right|^2 \, .
\label{NLOcollct}
\eeq
These results precisely correspond to the well-known, standard ones, as 
can be seen from \eq{treesqu} and \eq{AP0}. To avoid double counting, one 
must still subtract from the collinear counterterm the soft-collinear limit, which
emerges as expected from \eq{eikjetsq}. Pursuing the same approach at 
higher orders, starting at NNLO, we expect to develop a systematic 
viewpoint, which should allow us to take advantage of the well-organised 
structure of virtual corrections to simplify the interplay of real radiation 
counterterms. Clearly, what we have given here is just a set of definitions and 
prescriptions: they have the advantage of being valid to all perturbative 
orders and being universal across multi-particle massless scattering 
amplitudes; on the other hand, to build an actual subtraction algorithm, one 
still needs to work through the details of phase-space parametrisations,
momentum mappings, and, of course, the necessary integrations of 
counterterms over the unresolved phase spaces. Our approach to these
essential practical problems has been outlined in~\cite{Torrielli}.


\section{Perspectives}
\label{persp}

We have outlined an approach to the subtraction problem which starts from 
the well-known factorisation of virtual correction to multi-parton scattering 
amplitudes, and we have derived a completely general set of definitions 
for local subtraction counterterms, valid to all orders and for any massless 
gauge theory amplitude. We hope this approach will help in simplifying the 
structure of subtraction and the analytic integration of counterterms at NNLO 
and beyond. Work is in progress to construct a concrete subtraction algorithm 
incorporating these ideas into a practical and efficient framework, applicable 
to relevant LHC processes.


\section*{Acknowledgements}

\noindent The work of PT has received funding from the European Union Seventh 
Framework programme for research and innovation under the Marie Curie 
grant agreement N. 609402-2020 researchers: Train to Move (T2M).



\end{document}